\documentclass[twocolumn,preprintnumbers,amsmath,amssymb,superscriptaddress,aps]{revtex4}

\usepackage{graphicx}
\usepackage{dcolumn}
\usepackage{bm}

\begin{document}


\title{Nonlinear optical probe of tunable surface electrons on a topological insulator}

\author{D. Hsieh*}
\affiliation{Department of Physics, Massachusetts Institute of
Technology, Cambridge, MA 02139, USA}
\author{J. W. McIver*}
\affiliation{Department of Physics, Massachusetts Institute of
Technology, Cambridge, MA 02139, USA} 

\affiliation{Department of
Physics, Harvard University, Cambridge, MA 02138, USA}

\author{D. H. Torchinsky}
\affiliation{Department of Physics, Massachusetts Institute of
Technology, Cambridge, MA 02139, USA}
\author{D. R. Gardner}
\affiliation{Department of Physics, Massachusetts Institute of
Technology, Cambridge, MA 02139, USA}
\author{Y. S. Lee}
\affiliation{Department of Physics, Massachusetts Institute of
Technology, Cambridge, MA 02139, USA}
\author{N. Gedik}
\affiliation{Department of Physics, Massachusetts Institute of
Technology, Cambridge, MA 02139, USA}

\date{\today}


\begin{abstract}
We use ultrafast laser pulses to experimentally demonstrate that the second-order optical response of bulk single crystals of the topological insulator Bi$_2$Se$_3$ is sensitive to its surface electrons. By performing surface doping dependence measurements as a function of photon polarization and sample orientation we show that second harmonic generation can simultaneously probe both the surface crystalline structure and the surface charge of Bi$_2$Se$_3$. Furthermore, we find that second harmonic generation using circularly polarized photons reveals the time-reversal symmetry properties of the system and is surprisingly robust against surface charging, which makes it a promising tool for spectroscopic studies of topological surfaces and buried interfaces.
\end{abstract}

\maketitle

Electrons on the surface of a three-dimensional (3D) topological insulator \cite{Moore_Nature,RMP,Qi} are predicted to exhibit exotic electrical properties including protection against localization from non-magnetic impurities \cite{Moore_Nature,RMP} and photo-induced spin-polarized currents \cite{Hosur,Raghu}. When interfaced with ordinary materials, these surface states are predicted to evolve into new broken symmetry electronic phases with unconventional responses such as an anomalous half-integer quantum Hall effect \cite{RMP,Qi_QFT} or topological superconductivity \cite{RMP}. The recent discovery of 3D topological insulators in Bi$_{1-x}$Sb$_x$ \cite{Hsieh_Nature,Hsieh_Science}, Bi$_2$Se$_3$ and related materials \cite{Xia,Zhang,Hsieh_PRL} has generated great interest to measure the symmetry and electrical properties of a single surface or buried interface. However transport techniques have difficulty separating the response from different surfaces and the bulk \cite{Hadar,Checkelsky}, and require deposition of contacts and gates that may perturb the electronic structure. Optical techniques have been proposed as a contact free alternative that can be focused onto a single surface \cite{Qi_QFT,Tse,Maciejko,Raghu} or interface. However, experiments so far have been limited to the linear optical regime, which is again dominated by the bulk carrier response \cite{Sushkov,LaForge,Butch}.

A contact free probe that is potentially surface or interface sensitive is second-order nonlinear optical spectroscopy. In general, the electrical polarization of a material $P_i(\omega)$ has a dominant component linear in the driving optical field $E_j(\omega)$ as well as weaker components proportional to higher powers of $E_j(\omega)$, where $\omega$ is the optical frequency and the indices run through three spatial coordinates. Components that contain two powers of $E_j(\omega)$ are responsible for second harmonic generation (SHG). For electric dipole processes, the polarization $P_i = \chi^{(2)}_{ijk}E_jE_k$ is obtained from a third rank susceptibility tensor $\chi^{(2)}_{ijk}$ that vanishes under inversion symmetry. Therefore dipole induced SHG is forbidden in bulk crystals with inversion symmetry and is only allowed at surfaces or interfaces where inversion symmetry is necessarily broken \cite{Shen}. However, higher multipole bulk contributions to SHG are still allowed even in inversion symmetric systems, which can obscure the surface dipole contribution. Therefore it is unclear that SHG can be used as a surface or interface sensitive probe for a particular material. In this Letter, we generate SHG from Bi$_2$Se$_3$(111) by utilizing the high peak photon flux of ultrafast lasers. By performing surface doping dependence measurements as a function of photon polarization and sample orientation [Fig.1(a)] we show that SHG is surface sensitive and is a simultaneous probe of both the surface crystalline structure and the surface Fermi level of Bi$_2$Se$_3$. Furthermore, we find that circular dichroism SHG probes the time-reversal symmetry properties of the system and is robust against surface charging, which makes it a promising tool for spectroscopic studies of topological surfaces and buried interfaces.

Reflection SHG from the (111) surface of metallic Bi$_2$Se$_3$ bulk single crystals, which are bulk inversion symmetric (space group $D^{5}_{3d}$ \cite{Zhang}), was measured using 80 fs laser pulses derived from a Ti:sapphire oscillator. The laser pulses have a center wavelength of 795 nm (1.56 eV) and a repetition rate of 1.6 MHz, and are focused to a 20 $\mu$m spot on the sample at an incident angle of 45$^{\circ}$ with an incident power density of 0.63 kW/cm$^2$, which is below the damage threshold. Specularly reflected photons at the second harmonic energy (3.12 eV) were selectively isolated through spectral filtering and collected by calibrated photomultiplier tubes sensitive to 3.12 eV light.

\begin{figure*}
\includegraphics[scale=0.6,clip=true, viewport=-0.0in 0in 10.8in 4.1in]{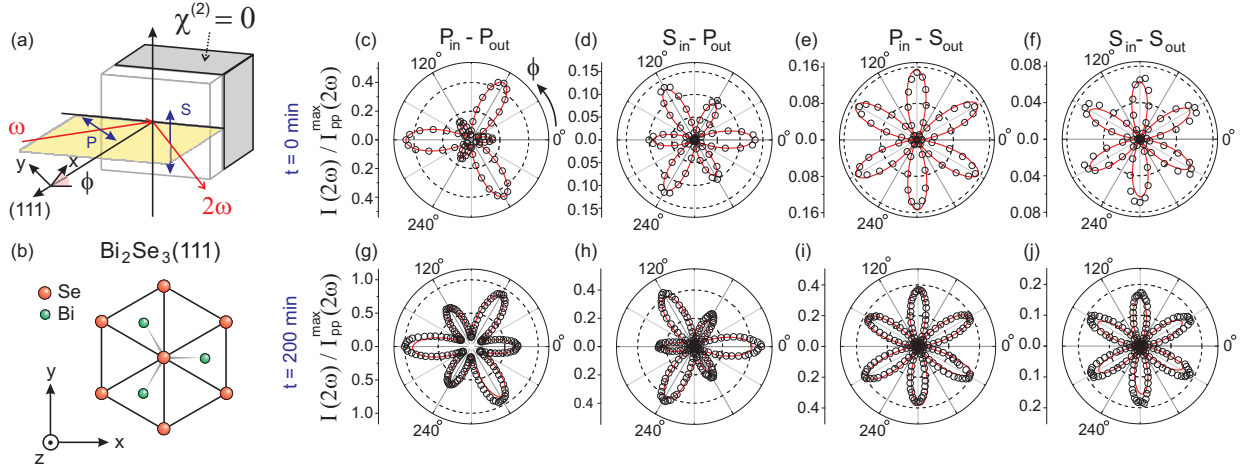}
\caption{\label{fig:linear} (a) Schematic of the SHG experimental geometry. Surface and bulk regions are colored white and gray respectively. (b) Crystal structure of the Bi$_2$Se$_3$ (111) surface showing the topmost Se layer and underlying Bi layer. (c) Normalized SHG intensity $I$(2$\omega$) from the (111) surface of Bi$_2$Se$_{3}$ measured as a function of azimuthal angle $\phi$ between the bisectrix (11$\bar{2}$) and the scattering plane. Measurements taken immediately after cleavage in $P_{in}$-$P_{out}$, (d) $S_{in}$-$P_{out}$, (e) $P_{in}$-$S_{out}$ and (f) $S_{in}$-$S_{out}$ incident and outgoing photon polarization geometries. Panels (g) to (j) shows analogous scans measured 200 minutes after sample cleavage. All data sets are normalized to the maximum intensity measured at $t = 200$ mins in $P_{in}$-$P_{out}$ geometry. The fewer data points in the early time scans is due to the need for faster data collection. Solid red lines are theoretical fits to Eqn.(\ref{eqn:polarization}) \cite{EPAPS}.}
\end{figure*}

We first demonstrate that our measurements are consistent with SHG from the (111) surface where the crystal symmetry is reduced from $D^{5}_{3d}$ to $C_{3v}$, which consists of an axis ($\hat{z}$) of 3-fold rotational symmetry and three planes of mirror symmetry [Fig.1(b)]. In general, the reflected surface SHG intensity $I(2\omega)$ is given by \cite{Mizrahi}

\begin{equation}
I(2\omega) = A \times |\hat{e}_i(2\omega) \chi^{(2)}_{ijk}
\hat{e}_j(\omega) \hat{e}_k(\omega)|^2 I(\omega)^2
\label{eqn:general}
\end{equation}
where $A$ is a constant determined by the experimental geometry, $\hat{e}$ is the polarization of the incoming or outgoing photons, $I(\omega)$ is the intensity of the incident beam and $\chi^{(2)}_{ijk}$ is a 27 component tensor. Application of $C_{3v}$ symmetry reduces $\chi^{(2)}_{ijk}$ to four non-zero independent
components $\chi_{xxx}$ = $-\chi_{xyy}$ = $-\chi_{yxy}$, $\chi_{zxx}$ = $\chi_{zyy}$, $\chi_{xxz}$ = $\chi_{yyz}$ and $\chi_{zzz}$. By controlling the relative orientation between the beam polarizations and crystal axes as shown in Fig.1(a), components describing the in-plane and out-of-plane electrical responses can be isolated via:

\begin{eqnarray}
I_{PP} (2\omega) &=& A \times |a^{(3)} - 0.025 \hspace{0.1cm} a^{(1)} \cos(3\phi)|^2 \nonumber \\
I_{SP} (2\omega) &=& A \times |a^{(2)} + 0.016 \hspace{0.1cm} a^{(1)} \cos(3\phi)|^2 \nonumber \\
I_{PS} (2\omega) &=& A \times |0.020 \hspace{0.1cm} a^{(1)} \sin(3\phi)|^2 \nonumber \\
I_{SS} (2\omega) &=& A \times |0.013 \hspace{0.1cm} a^{(1)}
\sin(3\phi)|^2 \label{eqn:polarization}
\end{eqnarray}
where the first and second sub-indices of $I(2\omega)$ denote the input and output polarizations respectively, $\phi$ is the angle between the scattering plane and the mirror ($xz$) plane of the crystal surface, $a^{(1)} \equiv \chi_{xxx}$ is an in-plane response and $a^{(2)} \equiv 0.061 \hspace{0.1cm} \chi_{zxx}$ and $a^{(3)}
\equiv -0.007 \hspace{0.1cm} \chi_{xxz} + 0.096 \hspace{0.1cm} \chi_{zxx} + 0.002 \hspace{0.1cm} \chi_{zzz}$ are out-of-plane responses \cite{EPAPS}.

Figure 1 shows the $\phi$ dependence of SHG intensity from the (111) surface of Bi$_2$Se$_{3}$ measured under different polarization geometries. The experiments were performed in air at room temperature and data were taken both immediately after and 200 minutes after the sample was cleaved. The time ($t$) dependence of the SHG patterns is related to surface charging as discussed later in the text. The SHG patterns exhibit a 6-fold or 3-fold rotational symmetry depending on whether $S$- or $P$-polarized output photons are selected respectively. Because $S$-polarized light only has in-plane electric-field components, it is only sensitive to the in-plane response $a^{(1)}$ likely originating from anharmonic polarizability of Se-Se bonds [Fig.1(b)], which have a 6-fold symmetric arrangement. On the other hand, because $P$-polarized light contains an electric-field component along $\hat{z}$, it is sensitive to out-of-plane responses $a^{(2)}$ and $a^{(3)}$ originating from the Se-Bi bonds, which have
a 3-fold symmetric arrangement. A fit of equations (\ref{eqn:polarization}) to the SHG patterns measured at a given time after cleavage using a single set of susceptibility tensor elements \cite{EPAPS} yields excellent agreement [Figs.1(c)-(f) \& Figs.1(g)-(j)], which shows that the data is consistent with surface SHG from Bi$_2$Se$_3$.

To prove that the surface contribution to SHG is dominant, we study its response to changes in the surface carrier concentration. It is known from angle-resolved photoemission spectroscopy (ARPES) \cite{Hsieh_Nature2} that after cleavage the surface electron concentration of Bi$_2$Se$_{3}$ increases monotonically over an hours long time scale before stabilizing. To study the effect of this surface electron doping on SHG we measure its time dependence immediately after cleavage. Figure 1(c)-(j) shows a clear change in the magnitude of all SHG patterns between $t = 0$ and $t = 200$ mins but no change in their rotational symmetry, which precludes a trigonal symmetry breaking atomic reconstruction. To investigate the possibility of surface atomic displacements that preserve overall trigonal symmetry such as a change of inter-layer distance, which would affect the in- and out-of-plane susceptibilities differently, we compare the complete time dependence of SHG intensities under different linear polarization geometries. Figure 2 shows that different combinations of susceptibility components all undergo the same monotonic increase by as much as 400\% between $t$ = 0 and $t$ = 50 mins following cleavage, after which they saturate to a value that remains constant out to at least 600 mins, a trend highly consistent with the evolution of the surface Fermi level observed using ARPES \cite{Hsieh_Nature2}. To rule out photo-induced effects, we performed time dependence experiments without exposing the sample to light for the first 100 mins after cleavage. In this case we find that the SHG intensities remain constant at the previous saturation values [insets Fig.2(c) \& (d)], which indicates that the initial rise in SHG intensity is purely a surface doping effect and not a photo-induced effect \cite{Qi_GaAs,Bloch,Mitchell}. Such large percentage changes in SHG intensity with surface electron doping suggest that SHG comes predominantly from electrons at the surface.

\begin{figure}
\includegraphics[scale=0.38,clip=true, viewport=0.0in 0in 9.5in 7.6in]{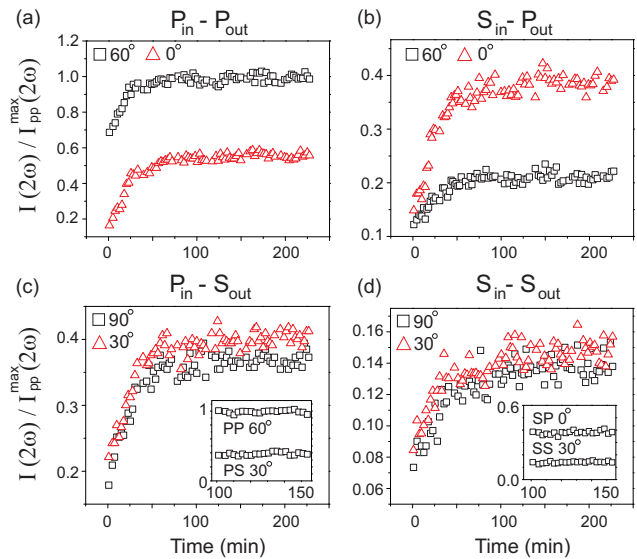}
\caption{\label{fig:time} Normalized SHG intensity along high symmetry directions of the (111) surface Brillouin zone of Bi$_2$Se$_{3}$ measured as a function of time after cleavage in air with (a) $P_{in}$-$P_{out}$, (b) $S_{in}$-$P_{out}$, (c) $P_{in}$-$S_{out}$ and (d) $S_{in}$-$S_{out}$ photon polarization geometries. Insets in panels (c) and (d) show the time evolution of peak intensities measured starting 100 minutes after cleavage in air, prior to which the sample was not exposed to laser light.}
\end{figure}

\begin{figure}
\includegraphics[scale=0.4,clip=true, viewport=0.0in 0in 9.3in 4.5in]{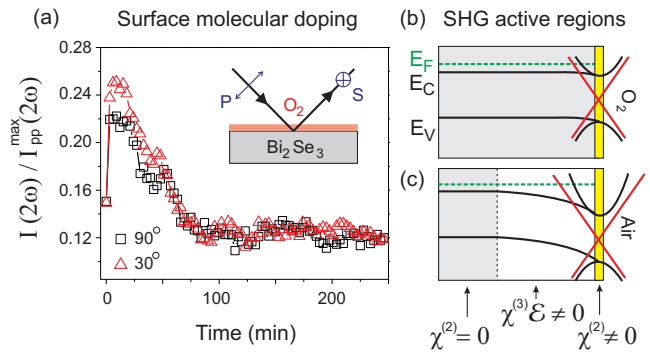}
\caption{\label{fig:Fits} (a) Selected time dependent peak intensity curves measured in $P_{in}$-$S_{out}$ polarization geometry after cleaving in O$_2$. A similar decrease was also observed in other polarization geometries. (b) Schematic of the energy evolution of the bulk conduction band minimum ($E_C$) and bulk valence band maximum ($E_V$) relative to the Fermi level ($E_F$) as a function of distance to the O$_2$ or (c) air covered surface. Typical Bi$_2$Se$_{3}$ samples show bulk conducting character ($E_F > E_C$) owing to Se vacancies \cite{Xia,Hsieh_Nature2}. The gray, white and yellow regions correspond to the bulk, accumulation layer and surface of Bi$_2$Se$_{3}$. Schematic of the bulk (black) and surface (red) band structure is drawn near the surface region.}
\end{figure}

Microscopically the increased surface carrier concentration affects SHG intensity through the perpendicular surface electric field $\vec{\mathcal{E}}$ that it creates. SHG is highly sensitive to a static electric field because it breaks inversion symmetry over the region that it penetrates, which is typically around 20 \AA\ for metallic Bi$_2$Se$_3$ \cite{Analytis}. Electric-field induced SHG is commonly observed in the context of metal/electrolyte interfaces, and is theoretically described by a third order process $P_i(2\omega)=\chi^{(3)}_{ijkl}\mathcal{E}_j(0)\hat{e}_k(\omega)\hat{e}_l(\omega)$ \cite{Bloch,Shen} that acts in addition to $\chi^{(2)}$. Because $\chi^{(3)}$ has the same symmetry constraints as $\chi^{(2)}$ \cite{EPAPS}, the overall symmetry of the SHG patterns must remain unchanged in the presence of $\vec{\mathcal{E}}$, with each tensor element in equation (\ref{eqn:polarization}) simply being enhanced by the addition of a $\chi^{(3)}\mathcal{E}$ tensor element. To demonstrate that SHG is sensitive to tuning the surface Fermi level we deposit O$_2$ on the surface, which has been shown to be an effective electron acceptor on Bi$_2$Se$_3$ (111) \cite{Chen2}. Figure 3 shows the time dependence of $I_{PS} (2\omega)$ after cleaving in an O$_2$ environment. We find that the SHG intensity exhibits an initial fast upturn reaching about 50\% of the saturation intensity value in air, followed by a gradual decrease back to its initial value. These results show that electron transfer from the Bi$_2$Se$_3$ surface to O$_2$ molecules takes place only after some finite $\vec{\mathcal{E}}$ has developed, and that O$_2$ can restore the surface back to an un-charged ($\vec{\mathcal{E}}$=0) state but cannot hole dope beyond this state. These behaviors are consistent with $\vec{\mathcal{E}}$ arising from the migration of negatively charged Se vacancies to the surface, which acts to lower the surface energy of the topmost Se layer. Assuming that charge transfer only takes place when an O$_2$ molecule is adsorbed at a surface Se vacancy site, no additional hole doping will occur once all vacancies are occupied. Such a mechanism naturally explains the slow time scale on which $\vec{\mathcal{E}}$ develops, as well as why the same behavior is observed regardless of whether the sample is cleaved in air or in ultra-high vacuum \cite{Hsieh_Nature2}.

\begin{figure}
\includegraphics[scale=0.38,clip=true, viewport=0.0in 0in 9.5in 7.8in]{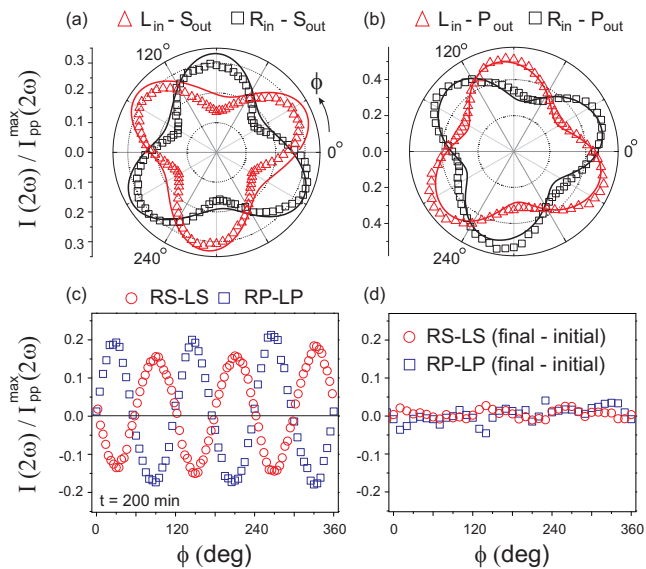}
\caption{\label{fig:circular} Normalized $\phi$-dependent SHG intensity patterns measured 200 minutes after cleavage in air under (a)
left-circular in ($L_{in}$) $S_{out}$ and right-circular in ($R_{in}$) $S_{out}$, and (b) $L_{in}$-$P_{out}$ and $R_{in}$-$P_{out}$ photon polarization geometries. Solid lines are theoretical fits (see text). (c) Circular dichroism ($I_{R}-I_{L}$) corresponding to data in panels (a) and (b). (d) Difference between circular dichroism measured at 200 minutes after cleavage and immediately after cleavage.}
\end{figure}

Having established an optical method to probe the spatial symmetry and Fermi level of the surface electrons of Bi$_2$Se$_3$, we consider how SHG can be used to monitor time-reversal symmetry at a topological insulator surface. It has been proposed that new time-reversal symmetry broken phases can be measured through the
differential index of refraction of right- ($R$) versus left- ($L$) circularly polarized light \cite{Qi_QFT,Tse,Maciejko}. However, it is known that the linear response of Bi$_2$Se$_3$ to circularly-polarized light is only sensitive to bulk electrons \cite{Sushkov}. In order to understand whether second harmonic versions of such experiments are feasible, it is necessary to measure the intrinsic second-order surface response of Bi$_2$Se$_3$ to circularly-polarized light.

Second harmonic circular dichroism (CD), the difference in SHG yield using incident $R$- versus $L$-polarized fundamental light, was measured from Bi$_2$Se$_{3}$ 200 minutes after cleavage in air. Figures 4(a) and (b) show that the SHG patterns obtained using $R$- and $L$-polarized incident light are clearly different for both $P$- and $S$-polarized output geometries, which are well described using equation (\ref{eqn:general}) with the same set of susceptibility values fitted to Figs.1(g) to (j). Although higher multipole bulk contributions to SHG, which are allowed in inversion symmetric crystals, may contribute to CD due to an interference with the surface dipole radiation \cite{Shen}, such bulk terms are known to be greatly suppressed relative to surface dipolar terms in the regime where the photon energy exceeds the bulk band gap \cite{Wang}. Our experiment is in this regime and we have experimentally shown that the surface contribution is indeed dominant (Figs 1-3). While CD is generally non-zero, figure 4(c) shows that it vanishes when $\phi$ is an integer multiple of 60$^{\circ}$, angles where the scattering plane coincides with a mirror plane of the (111) surface. Such zeroes are protected by mirror symmetry because $R$- and $L$-polarized light transform into one another under mirror reflection about the scattering plane. Because a magnetization can break mirror symmetry, measuring departures from zero in CD along these specific values of $\phi$ can be a sensitive probe of time-reversal symmetry breaking on the surface of Bi$_2$Se$_3$. Remarkably, we find that CD at a general $\phi$ is insensitive to surface charging, as evidenced by the lack of measurable change as a function of time after cleavage [Fig.4(d)]. This suggests that sensitive searches for time-reversal symmetry breaking induced CD may be carried out without the need for careful control of surface charging, which is an important and robust way of studying the physics of surface doped topological insulators or buried interfaces between topological insulators and ordinary materials \cite{Moore_Nature,RMP}.

The realization of surface second harmonic generation from Bi$_2$Se$_3$ and the demonstration of its sensitivity to the surface Fermi level, crystal symmetry and time-reversal symmetry provides a novel contact-free probe of the properties of a single surface of a topological insulator. This will make it possible to probe the structural and electronic properties of a buried topological insulator interface, and to search for exotic broken symmetry interface phases. The use of ultrashort laser pulses will also make it possible to selectively study time-resolved non-equilibrium dynamics of a surface or interface following an excitation, which for example can be applied to probe photo-induced spin-polarized currents.

\vspace{0.5cm}

\noindent \textbf{Acknowledgements.} This work was supported by
Department of Energy award number DE-FG02-08ER46521. D.H. is
supported through a Pappalardo Postdoctoral Fellowship. J.W.M is supported by an NSF Graduate Research Fellowship. 

\vspace{0.5cm}

\noindent \textbf{Author contributions.} *D.H. and J.W.M.
contributed equally to this work.\\

\newpage

\vspace{1cm}


\vspace{1cm}


\vspace{1cm}


\vspace{1cm}

\vspace{3.3cm}


\vspace{0.5cm}


\vspace{0.5cm}



\end{document}